\documentclass[preprint,showpacs,preprintnumbers]{revtex4}     
\usepackage{graphicx}
\usepackage{dcolumn}
\usepackage{bm}
\usepackage{ifthen}
\usepackage{booktabs}
\usepackage{amsmath}

\begin{document}

\title{Measurement-induced quantum entanglement recovery}
\author{Xiao-Ye Xu, Jin-Shi Xu, Chuan-Feng Li$\footnote{
email: cfli@ustc.edu.cn}$, and Guang-Can Guo} \affiliation{Key
Laboratory of Quantum Information University of Science and
Technology of China, CAS, Hefei, 230026, People's Republic of China}

\begin{abstract}

By using photon pairs created in parametric down conversion, we
report on an experiment, which demonstrates that measurement can
recover the quantum entanglement of two qubit system in a pure
dephasing environment. The concurrence of the final state with and
without measurement are compared and analyzed. Furthermore, we
verify that recovered states can still violate Bell's inequality,
that is, to say, such recovered states exhibit nonlocality. In the
context of quantum entanglement, sudden death and rebirth provide
clear evidence, which verifies that entanglement dynamics of the
system is sensitive not only to its environment, but also on its
initial state.

\end{abstract}

\maketitle


Quantum entanglement, as a unique feature without a classical
counterpart of many-body system, has been instrumental in studying
fundamental aspects of quantum physics \cite{Zeilinger} as well as
being central in practical applications in the areas of quantum
computation and quantum cryptography
\cite{Nielsen,Raussendorf,Brien,Ekert}. Today, sources of entangled
states can be prepared in various kinds of physical systems
\cite{Ladd}. Entanglement, even within multi-particle and
multi-dimensional systems \cite{Rossi,Lu}, can be implemented,
although, these are flimsy and are subject to unavoidable
degradation, which is caused by interactions with their environments
\cite{Zurek03,Mintert}. To have any practical value in quantum
computation and communication, long distance nonlocality and
extended coherence storage and rebirth \cite{Yu09} of entangled
states have become important focal points of research around the
world. The key issue behind solving these problems is in determining
the dynamical behavior of entanglement within the system's
environment, something, which to date, has not been well understood.
Commendably, a factorization law, which describes the entanglement
dynamics under a one-side noisy channel has recently been proposed
 \cite{Konrad} and has subsequently been verified in two independent
experiments \cite{Farias}. Moreover, it has been discovered that
entanglement evolution is not only related to environmental factors
but is also sensitive to its initial state \cite{Roszak,Yu06}.

Quantum measurement, that feature, which distinguishes the quantum
from the classical regimes \cite{Zurek91}, is often interpreted
within the orthogonal projection model given by Von Neumann
\cite{Neumann}, but has been reexpressed in the past 30 years or so,
more and more within the framework of quantum decoherence theory
\cite{Zurek81} (for reviews see \cite{Zurek03,Zurek91}). In that
setting, a complete quantum measurement is divided into two stages:
The first stage corresponds to the entanglement of the
information-carrying qubit of the measured quantum system with the
record bit of the measurement apparatus; the second stage
corresponds to the decoherence of the detector-system combination,
which occurs in an uncontrollable environment \cite{Zurek91}. The
latter will covert the density matrix that describes the combination
into diagonal form, which physically represents the spreadings of
the quantum information contained in the combined system into the
uncontrollable environment and essentially turns the quantum
measurement into a classical one. Because of the uncontrollability
and unavoidability of environment-induced decoherence that features
prominently in this second stage, it is impossible to recover
quantum information once the quantum measurement has been completed.
For example, if a conventional experiment to measure the
polarization of a single photon has completed, to retrieve that
qubit of quantum information is as difficult as extracting it from
the single photon detector and the observer's brain. Fortunately,
quantum measurement need not be so catastrophic and can be
implemented step by step, even partially \cite{Elitzur}. It has been
pointed out that, within the first stage of quantum measurement,
recovery can be brought about by the coherence of a single qubit
that has dissipated into a non-Markovian environment \cite{Xu09}.

On the basis of this work by Xu \emph{et al} \emph{Xu} we
experimentally prove that subsequent quantum measurement, which
erases \cite{Scully} the path information introduced in the previous
measurement can recover the entanglement that has been degraded in a
non-Markovian environment and that even rebirth of entanglement
after entanglement sudden death (ESD) \cite{Yu09} can be effected.
In this paper, we describe how we can change the entanglement
evolution by using some specified operations on the state before or
during the interaction, more precisely, two sequential measurements
can recover the entanglement and, further more, preserve it.


The experimental setup is shown schematically in Figure\,1. Two
0.5\,mm thick beta-barium-borate (BBO) crystals, cut at
29.18${^\circ}$ for type-\uppercase\expandafter{\romannumeral1}
phase matching and aligned so their optical axes are perpendicular
to each other, are pumped by using focused ultraviolet (UV) pulses
polarized at 45$^{\circ}$, which are frequency doubled from a
Ti:sapphire laser with the center wavelength mode locked at 800\,nm
(with 130\,fs pulse width and a 76\,MHz repetition rate). Degenerate
polarization-entangled photon pairs at 800\,nm are generated by
spontaneous parametric down conversion (SPDC) at a 3$^{\circ}$ angle
with the pump beam \cite{Kwiat99}. By compensating the time
difference between $H$- and $V$-polarized components with
birefringent elements (LiNbO$_{3}$ and YVO$_{4}$), one of the
maximal polarization-entangled states, the well-known Bell state
\cite{Bell}, can be produced with high fidelity. This initial state
can be mathematically written as
\begin{eqnarray}
|\phi\rangle = \frac{1}{\sqrt{2}}(|HH\rangle+|VV\rangle),
\end{eqnarray}
where \emph{H} and \emph{V} represent the horizontal and vertical
polarizations, respectively, while the two elements in the Dirac ket
label the photons states, the left in path $a$ and the right in path
$b$.

\begin{figure}
\begin{center}
\includegraphics[width=5in]{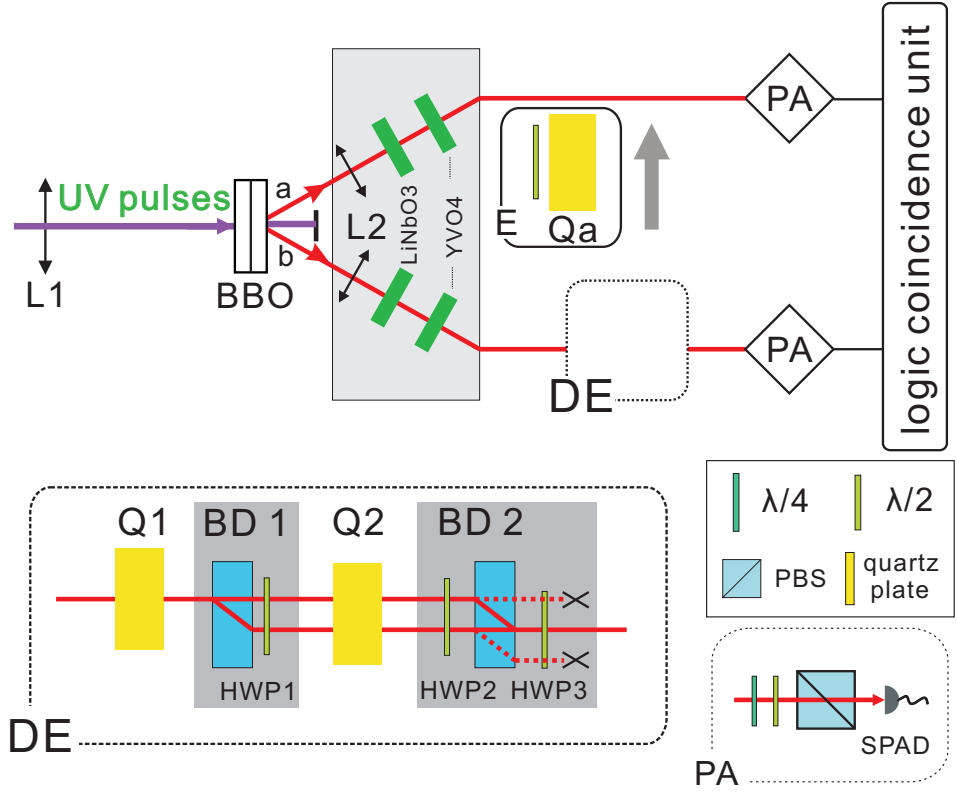}
\end{center}
\caption{(color online) Scheme of the experimental setup. DE,
decoherence evolution denoted by a dashed pane; Measurement
apparatus (M) is denoted by two gray boxes. PBS, polarizing beam
splitter; L1 and L2, lens; PA, polarization analyzer; SPAD, single
photon detector; BD1 and BD2, beam displacing prism; The solid pane
E is inserted in path \emph{a} to prepare the partially entangled
state. The parameter of these elements are provided in the text.}
\label{fig:setup}
\end{figure}

Decoherence due to the environment is simulated by controllable
birefringent elements, which can couple the photon's frequency with
its polarization \cite{Kwiat00,Berglund}. In our experiment, we use
quartz plates Q1 with thickness $L_1$ and Q2 with thickness $L_2$,
to simulate this decoherence aspect. The optical axes of the plates
are both horizontally set.

At the beginning, we consider the evolution of the state given by
equation\,(1) in such an environment, which assume, for simplicity,
that the photon's frequency distribution is a $\delta$ function. The
final state of the two photons after the photon in mode $b$ passes
through Q1 and Q2 takes the following form
\begin{eqnarray}
|\phi_{1}\rangle =
\frac{1}{\sqrt{2}}(|HH\rangle+e^{i\alpha\omega_{b}}|VV\rangle),
\end{eqnarray}
where the parameter $\alpha$ is proportional to $L(=L_1+L_2)$ and
$\omega_{b}$ represents the frequency of the photon in path $b$. In
our experiment, $\alpha = L\Delta n/c$ where $c$ is the velocity of
light in vacuo and $\Delta n = n_{o}-n_{e}$ represents the
difference between refractive indices of ordinary $(n_{o})$ and
extraordinary $(n_{e})$ light. Therefore, a deterministic relative
phase is introduced between $|HH\rangle$ and $|VV\rangle$ for a
single frequency, which differs for different frequencies. According
to the decoherence mode \cite{Kwiat00,Berglund}, the environment is
actually composed of a photon's frequencies that are coupled to the
information carriers (viz. polarization of photons) by means of
birefringent elements. For photons with a frequency distribution
$f(\omega)$, the overall state should equal the integral over the
frequency distribution, which forms a less correlated state, which
essentially destroys the coherence of the qubits \cite{Kwiat00}.
Therefore, the final state in Eq.\,(2) should be replaced by the
following reduced density operator \cite{Berglund}
\begin{eqnarray}
\hat{\rho}_{1} = \frac{1}{2}(|HH\rangle\langle HH| +
|VV\rangle\langle VV| + k_{b}^{*}|HH\rangle\langle VV| +
k_{b}|VV\rangle\langle HH|),
\end{eqnarray}
where the photon's frequency distribution in path $b$ is normalized
as $\int f(\omega_{b})d\omega_{b}=1$ and the nondiagonal coefficient
$k_{b} = \int f(\omega_{b})e^{i\alpha\omega_{b}}d\omega_{b}$
represents the decoherence parameter.

Actually, for a decoherence environment composed of photon's
frequency, the decoherence parameter $k_{b}$ is related to the
frequency distribution function. In our experiment, this is taken to
be a Gaussian function, that is, $f(\omega_{b}) =
\frac{2}{\sqrt{\pi}\sigma}\exp(-\frac{4(\omega_{b}-\omega_{0})^2}{\sigma^2})$,
which is determined by the interference filters (IF) placed in front
of the single photon detectors. The parameter $\omega_{0}$ is the
central frequency, and $\sigma$ is the bandwidth. By working out the
integral, we obtain  $k_{b} =
\exp(-\alpha^2\sigma^2/16+i\alpha\omega_{0})$.

For a quantitative analysis of the entanglement evolution in the
experiment, a parameter, which represents the degree of
entanglement, should be introduced. The concurrence \cite{Wootters},
which is widely used in studying two qubit states, is defined as
\begin{eqnarray}
C(\hat{\rho}) =
\max\{0,\sqrt{\lambda_{1}}-\sqrt{\lambda_{2}}-\sqrt{\lambda_{3}}-\sqrt{\lambda_{4}}\},
\end{eqnarray}
where $\hat{\rho}$ is the density matrix of a two-qubit state in the
canonical basis $\{|HH\rangle,|HV\rangle,|VH\rangle,|VV\rangle\}$;
$\lambda_{i} (i = 0,\ldots, 4)$ are the eigenvalues in decreasing
order of the Hermitian matrix
$\hat{\rho}(\hat{\sigma}_{y}\otimes\hat{\sigma}_{y})\hat{\rho}^{*}(\hat{\sigma}_{y}\otimes\hat{\sigma}_{y})$
with $\hat{\rho}^{*}$, which corresponds to the complex conjugate of
$\hat{\rho}$. According to Eq.\,(3), for an initial Bell state
input, we obtain $C(\hat{\rho}_{1}) = |k_{b}| =
\exp(-\alpha^2\sigma^2/8)$. Thus, concurrence degrades exponentially
and approaches zero as Q1 and Q2 become thicker, which means the
final state, after sufficient interaction time, evolves into the
maximally mixed state without any remaining entanglement [solid line
in Fig.\,2(A)]. However, for some other frequency distribution, some
unusual phenomena will arise, for example, entanglement collapse and
revival \cite{Xu10}.

Then, we consider the case with a measurement apparatus (M), which
comprises two beam displacing prisms (BD1 and BD2) with
horizontally-set optical axes, half-wave plate 1 (HWP1) with optical
axes set at 22.5$^\circ$, which implements the Hadamard operation,
HWP2 with optical axes set at -22.5$^\circ$, which implements
$|H\rangle\rightarrow|+\rangle,|V\rangle\rightarrow-|-\rangle$
(where,$|+\rangle= (|H\rangle+|V\rangle)/\sqrt{2}$, $|-\rangle=
(|H\rangle-|V\rangle)/\sqrt{2}$), and HWP3 with perpendicularly set
optical axes to implement the bit-flip operation. BD1 measures
photon's polarization in H/V basis and introduces the path
information as a probe bit. Before this measurement is completed,
the second decoherence environment is inserted, and then, the path
information is erased by BD2, which realizes a postselection of the
recovered state. The output state after passing through Q1, BD1,
HWP1 and Q2, can be written as
\begin{eqnarray}
(\frac{1}{2}|H\rangle(|H\rangle +
e^{i\alpha_{2}\omega_{b}}|V\rangle))_{\texttt{\uppercase\expandafter{\romannumeral1}}}
+(\frac{1}{2}e^{i\alpha_{1}\omega_{b}}|V\rangle(|H\rangle -
e^{i\alpha_{2}\omega_{b}}|V\rangle))_{\texttt{\uppercase\expandafter{\romannumeral2}}}
\end{eqnarray}
where $\alpha_{1} = L_{1}\Delta n/c$ and $\alpha_{2} = L_{2}\Delta
n/c$, subscripts \uppercase\expandafter{\romannumeral1} and
\uppercase\expandafter{\romannumeral2} denote the upper and lower
paths between the two BDs, respectively. By subsequently erasing
path information introduced in the previous measuring operation by
HWP2, BD2 and HWP3, we obtain the final state represented as
\begin{eqnarray}
|\phi_2\rangle =
\frac{1}{2}(1+e^{i\alpha_{2}\omega_{B}})(|HH\rangle+e^{i\alpha_{1}\omega_{B}}|VV\rangle).
\end{eqnarray}
Similar to the above treatment, the reduced density matrix of this
final state is written as
\begin{eqnarray}
\hat{\rho}_{2} = \frac{1}{2}(|HH\rangle\langle HH| +
|VV\rangle\langle VV| + k_{b}'^{*}|HH\rangle\langle VV| +
k_{b}'|VV\rangle\langle HH|),
\end{eqnarray}
where
\begin{eqnarray}
k_{b}' &=&
\frac{\exp(i\alpha_{1}\omega_{0})}{2[1+\cos(\alpha_{2}\omega_{0})\exp(-\alpha_{2}^{2}\sigma^{2}/16)]}
\{\exp[-(\alpha_{1}+\alpha_{2})^{2}\sigma^2+i\alpha_{2}\omega_{0}]\nonumber\\
&+&\exp[-(\alpha_{1}-\alpha_{2})^{2}\sigma^2-i\alpha_{2}\omega_{0}]
+2\exp[-\alpha_{1}^{2}\sigma^{2}/16)] \}\nonumber.
\end{eqnarray}
According to Eq.\,(4), the concurrence is $C(\hat{\rho}_{2}) =
|k_{b}'|$. From the complicated form of $k_{b}'$, here, the
entanglement evolution is not as simple as that in the previous
case, and it sensitive to the phase. Numerical analysis shows there
will be entanglement recovery with increases in $L_2$, as $L_1$
remains fixed. The concurrence oscillates if $L_2$ is sufficiently
thin, while the amplitude narrows to zero if $L_2$ is thick enough.
Because the maximal recovery point lies within the envelope composed
of the integral $L_2$ and the zero point of the phase is extremely
hard to determine, here, we only consider the integral length of the
quartz plates [solid line in Fig.\,2(B) and Fig.\,2(C)]. Although
this is sufficient for studying entanglement recovery, under this
consideration, the probability of success decreases exponentially
from 1 to 0.5. More significantly, the entanglement of the final
state will remain unchanged with $L_2$ increasing when $L_2$ is
large enough, and no matter how thick $L_1$ is, that is, to say, no
matter how less entanglement remains, the entanglement between two
particles can be recovered to a maximal expectation value 0.5 at
$L_2=L_1$.

In fact, the dynamics of entanglement in bipartite quantum systems
not only is sensitive to their environment, but also is sensitive to
their initial state \cite{Roszak,Yu06}. In Ref \cite{Xu10},
entanglement collapse and revival occur when the input biphoton is
prepared in the form of a Werner state \cite{Werner} with a spectrum
discretized within a Gaussian envelope. There will be no
entanglement revival if the spectrum takes the Gaussian form in that
experiment. However, a revival of the same initial state with a
Gaussian spectrum can occur by inserting M in this experiment. This
is explained as follows.

Applying a Hadamard operation on the photon state in mode $a$ of the
maximal entangled state $|\phi\rangle$ and by allowing it to pass
though a dephasing environment at $H/V$ bases, we get the state
\begin{eqnarray}
|\phi_{3}\rangle =
\frac{1}{2}(|HH\rangle+|HV\rangle+e^{i\alpha_{a}\omega_{a}}|VH\rangle-e^{i\alpha_{a}\omega_{a}}|VV\rangle).
\end{eqnarray}
By integrating $|\phi_{3}\rangle\langle\phi_{3}|$ over all
frequencies of the photon in mode $a$, the partially entangled input
state can be mathematically expressed in the following density
matrix
\begin{eqnarray}
\hat{\rho}_{0} = \frac{1}{4}
\begin{pmatrix}
1        & 1        & k_{a}^{*}   & -k_{a}^{*} \\
1        & 1        & k_{a}^{*}   & -k_{a}^{*} \\
k_{a}    & k_{a}    & 1           & -1 \\
-k_{a}   & -k_{a}   & -1          & 1
\end{pmatrix},
\end{eqnarray}
where $k_{a} = \int
g(\omega_{a})\exp(i\alpha_{a}\omega_{a})d\omega_{a}$ is the
decoherence parameter in path $a$. The photon in mode $b$ then
passes through the decoherence environment with a measuring
apparatus; the final state in the single frequency case can be
written as
\begin{eqnarray}
|\phi_{3}'\rangle =
\frac{1+e^{i\alpha_2\omega_b}}{4}[|HH\rangle+e^{i\alpha_{1}\omega_{b}}|HV\rangle
+e^{i\alpha_{a}\omega_{a}}|VH\rangle-e^{i(\alpha_{a}\omega_{a}+\alpha_{1}\omega_{b})}|VV\rangle].
\end{eqnarray}
The band width of the interference filters used in our experiment is
so narrow that we can integrate $|\phi_{3}'\rangle\langle\phi_{3}'|$
over the frequencies of the photon in mode $a$ and $b$ separately
\cite{Ou}. Therefore, if the photon frequencies are considered to
have Gaussian distribution profiles, the density matrix of the final
state is written as
\begin{eqnarray}
\hat{\rho}_{3} = \frac{1}{4}
\begin{pmatrix}
1        & k_{b}'^{*}        & k_{a}^{*}   & -k_{a}^{*}k_{b}'^{*} \\
k_{b}'        & 1        & k_{a}^{*}k_{b}'   & -k_{a}^{*} \\
k_{a}    & k_{a}k_{b}'^{*}    & 1           & -k_{b}'^{*} \\
-k_{a}k_{b}'   & -k_{a}   & -k_{b}'          & 1
\end{pmatrix}.
\end{eqnarray}
According to Eq.\,(4), we obtain a concurrence of $C(\hat{\rho}_{3})
= \max\{0,(k_{a}+k_{b}'+k_{a}k_{b}'-1)/2\}$. For nonnegative
$(k_{a}+k_{b}'+k_{a}k_{b}'-1)/2$ with different $k_{b}'$,
entanglement, sudden death and re-birth will occur if the
decoherence time in Q1 and Q2 are changed [solid line in Fig.\,2(D)
with just integral lengths of quartz plates]. No matter how thick Q1
is, entanglement between the two particles is found to indicate a
full rebirth to an identical maximal value at $L_2=L_1$. That is, to
say, full recovery with analogous levels can be achieved regardless
of the time duration taken for the ESD.

\begin{figure*}
\includegraphics[width=2.8in]{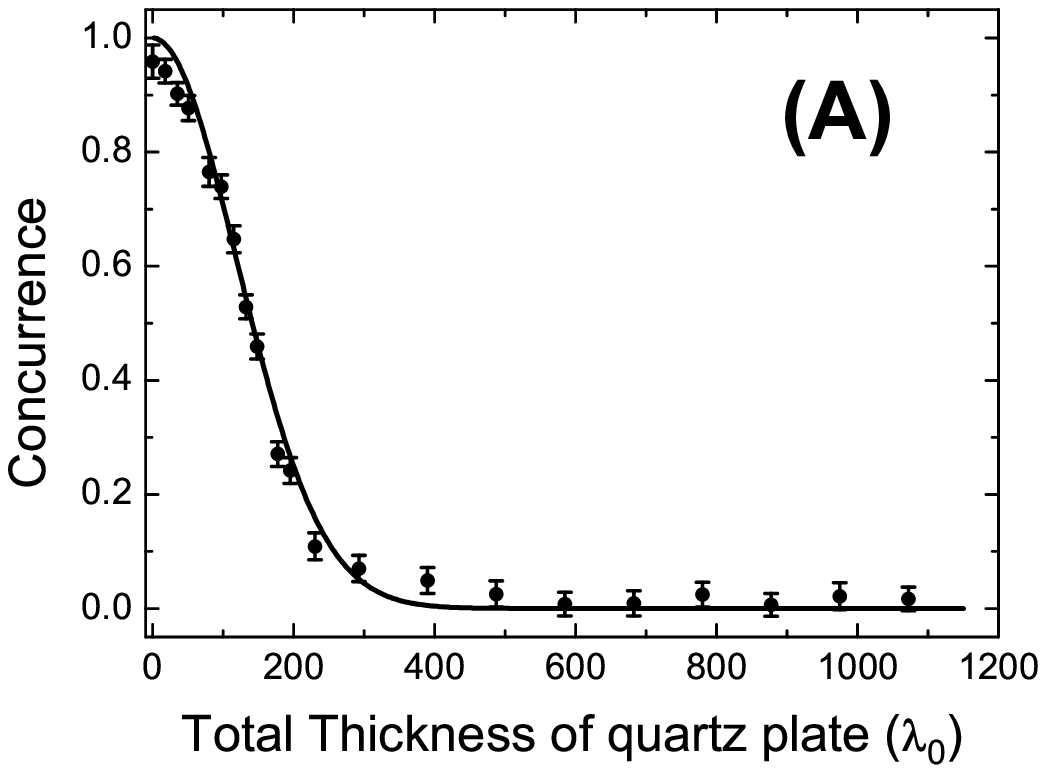}
\includegraphics[width=2.8in]{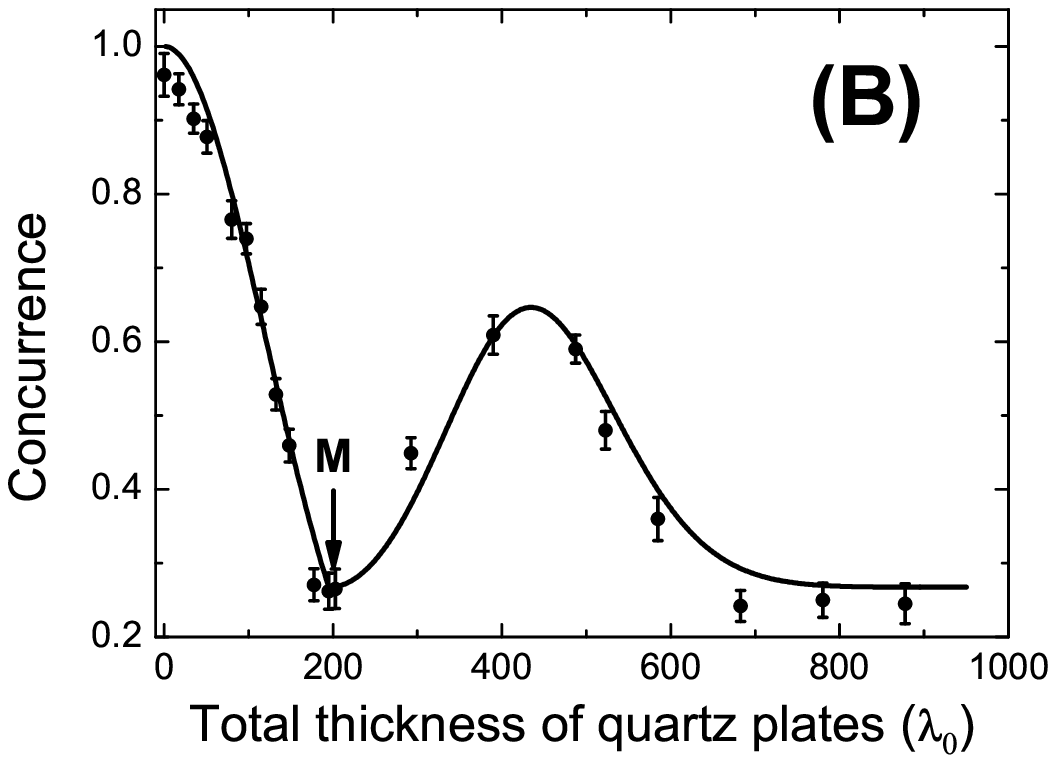}
\includegraphics[width=2.8in]{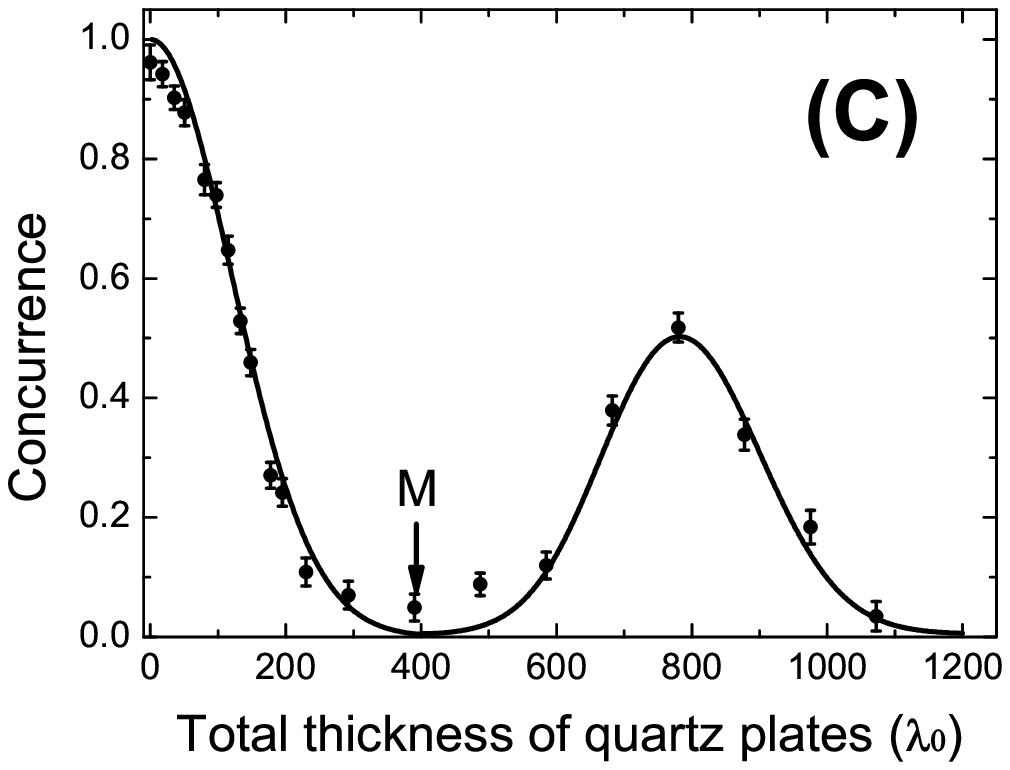}
\includegraphics[width=2.8in]{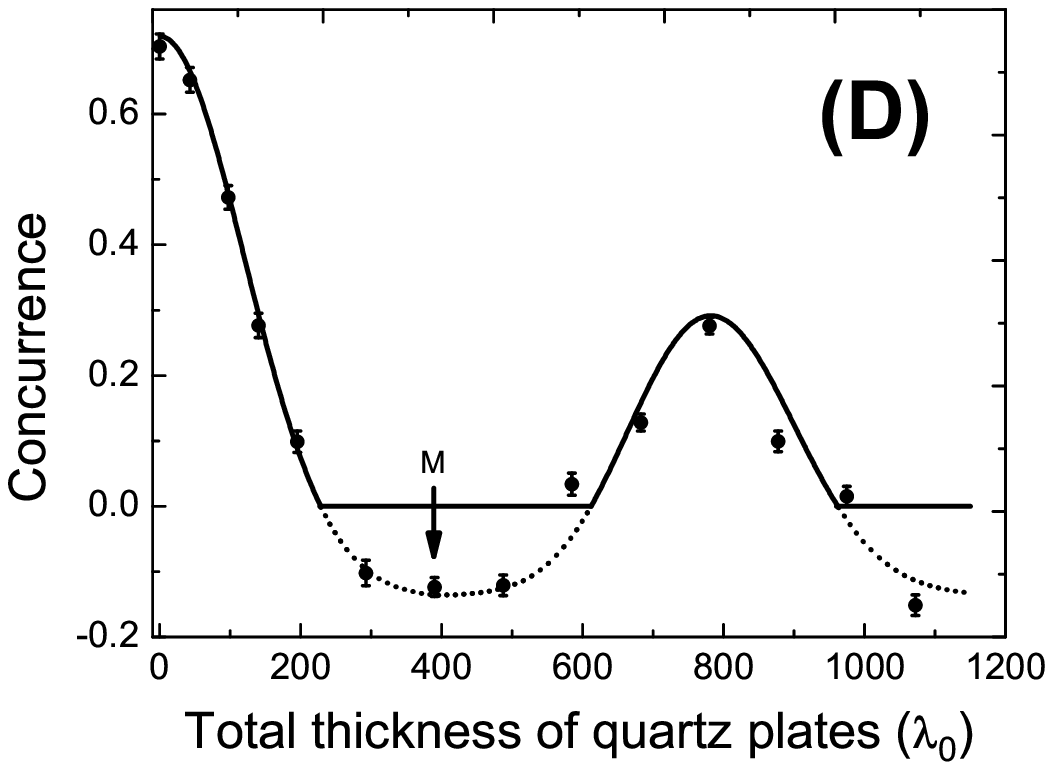}
\caption{ Major results (dots) for entanglement dynamics in the
experiment. The solid lines are theoretical predictions of
concurrence. Two error types are considered: the shot noise error in
the measured coincidence counts and the uncertainty in the settings
of the angles of the wave plates used to perform the tomography
\cite{James}. (a) represents the entanglement evolution of a maximal
entangled state in a pure dephasing environment without M; (b) and
(c) with M inserted at $L_1=195$ and $L_1=390$, respectively; (d)
represents the entanglement evolution of the partially entangled
state in a dephasing environment with M inserted at $L1=390$.
$\lambda_0=800$\,nm.} \label{fig:result}
\end{figure*}

The experimental results are shown in Fig.\,2. The characteristics
of the IFs used before the single photon detectors are of bandwidth
3\,nm and coatings at 800\,nm. We can treat $\Delta n=0.01$ for
small frequency distributions. The maximally entangled state is
prepared with a concurrence of 0.962\,$\pm$\,0.029. In Fig.\,2(a),
the concurrence degrades exponentially and gradually tends to zero,
which obeys the half-life law. Because there is no phase
sensitivity, experimental results (dots) agreed well with theory
within the error range. In Fig.\,2(b), M is inserted at the point
$L_1=195\lambda_0$ where the concurrence is 0.262\,$\pm$\,0.024. The
maximally recovered concurrence in the experiment is
0.609\,$\pm$\,0.026 at $L_2=195\lambda_0$, and the concurrence is
unchanged within the error range when $L_2$ exceeds 683$\lambda_0$.
In Fig.\,2(c), the point at which M is inserted is
$L_2=390\lambda_0$ where the concurrence tends to zero, that is, to
say, there are few entanglement at this point. Incidentally, the
maximally recovered entanglement measured in the experiment is
0.518\,$\pm$\,0.025 at about $L_2=780\lambda_0$, which agrees well
with theoretical predictions. In Fig.\,2(d), the partially entangled
input state with concurrence 0.704\,$\pm$\,0.019 is prepared by
inserting an HWP with optical axes set at 22.5$^\circ$ and quartz
plates of thickness 98$\lambda_0$ with horizontally-set optical axes
 in mode \emph{a}. We insert M at $L_1=390\lambda_0$ at which ESD has occurred, and
there is no entanglement; entanglement rebirth occurs at about
$L_2=585\lambda_0$ and then the entanglement collapses at about
$L_2=975\lambda_0$ again. The concurrence is corrected to zero
according to Eq.\,(4) when its measured value is negative.
Entanglement can be reborn at a maximal value 0.276\,$\pm$\,0.013 in
the experiment.

Nonlocality, as a particular characteristic of quantum mechanics,
has changed the viewpoint and methods in understanding nature at its
fundamental level. Since it can be studied by the well known Bell
inequality \cite{Bell}, a quantum state with nonlocal correlations
could be a very useful feature to exploit in future quantum
technologies \cite{Gisin}. In our experiment, the maximally
recovered entangled state in Fig.\,2(b) and Fig.\,2(c) can be proven
to be nonlocal by the more convenient Clauser-Horne-Shimony-Holt
(CHSH) inequality \cite{Clauser}, $S\leq2$ for any local realistic
theory, where
\begin{eqnarray}
S =
E(\theta_1,\theta_2)+E(\theta_1,\theta_2')+E(\theta_1',\theta_2)-E(\theta_1',\theta_2')
\end{eqnarray}
with
\begin{eqnarray}
E(\theta_1,\theta_2)
=\frac{C(\theta_1,\theta_2)+C(\theta_1^\perp,\theta_2^\perp)-C(\theta_1,\theta_2^\perp)-
C(\theta_1^\perp,\theta_2)}{C(\theta_1,\theta_2)+C(\theta_1^\perp,\theta_2^\perp)+C(\theta_1,\theta_2^\perp)+
C(\theta_1^\perp,\theta_2)}.\nonumber
\end{eqnarray}
Here, $\theta_i(\theta_i'),i=1,2$ represent the linear polarization
setting in path \emph{a} and path \emph{b} separately and
$\theta_i^\perp=\theta_i+90^\circ,i=1,2$. By calculating the maximal
value of \emph{S} from the measured density matrix of the maximally
recovered state, we get $(\theta_1=-15^\circ, \theta_1'=21^\circ,
\theta_2=86^\circ, \theta_2'=-52^\circ)$ in Fig.\,2(B) and
$(\theta_1=-82^\circ, \theta_1'=66^\circ, \theta_2=-4^\circ,
\theta_2'=28^\circ)$ in Fig.\,2(C). Accordingly, the measured values
of \emph{S} are 2.336\,$\pm$\,0.003 and 2.210\,$\pm$\,0.003, which
violate the local realism limit 2 by over 104 and 64 standard
deviations, respectively.

The extraordinary phenomenon of entanglement recovery induced by a
measurement, can be understood in the quantum framework of a partial
measurement and reversal \cite{Elitzur}. BD1 measures the photon's
polarization in H/V basis. Partial measurement in H/V basis in the
upper path and a reversal operation with the same strength
\cite{caption} in the lower path is implemented by BD2. The photons
in the two dark ports of BD2 are abandoned, this can be considered
as a non-response of the detector to the coincidence detection.
Because partial measurement and reversal with the same intensity can
restore the initial entanglement \cite{Elitzur} and the phase
difference introduced by $L_2$ only changes the strength, we believe
that entanglement is preserved with the increasing $L_2$ while the
recovery is an optical spin-echo effect introduced by HWP1.

To summarize, we have report on an experiment, which shows that
entanglement can be recovered by a process of measurement followed
by quantum eraser in a non-Markovian environment. Simultaneously,
the maximally recovered states can be verified to violate the CHSH
inequality with high standard deviations, which confirms theirs
quantum character. This result can be used to eliminate the
influence of dephasing. Another encouraging aspect is that, even if
the state has been thoroughly disentangled, that is, to say, ESD has
occurred, the entanglement can still be revived from non-Markovian
environments regardless of decoherence time durations. This can be
used to implement controllable recovery of entanglement.
Entanglement dynamics considering other coding and decoding protocol
can be studied in future work.

This work was supported by National Fundamental Research Program and
the National Natural Science Foundation of China (Grant No.60621064
and 10874162).


\begin{thebibliography}{99}
\bibitem{caption} The stength will attenuate from 1 to 0.5 with
$L_2$ increasing while keeping the equivalence between partial
measurement and erasure.
\end{thebibliography}
\end{document}